# Specially Coupled Dark Energy in the Oscillating FRW Cosmology

A.V. Timoshkin*

*Tomsk State Pedagogical University, Tomsk, Russia*

**Abstract:** We consider a four-dimensional flat-space Friedman universe, which is filled with two interacting ideal fluids (the coupling of dark energy with dark matter of special form). The gravitational equations of motion are solved. It is shown that in some cases there appears a periodic universe with finite-time cosmological singularities and also the universe becomes static in the remote future.

## 1. INTRODUCTION

In the present work we investigate a phenomenological model of the coupling of dark energy with dark matter of special form (for a review, see [1]), where there is a dark energy with a non-linear inhomogeneous equation of state and a dark matter with a linear homogeneous equation of state (for a review of similar models in modified gravity, see reference [2]). There is a lot of interest in the study of a nature of the coupling of dark matter with dark energy responsible for the acceleration of cosmic expansion. Among the different possible models that have been considered in the literature there exists a model in which a dark energy and a dark matter are described by the ideal fluid with an unusual equation of state. Very general dark-fluid models with an inhomogeneous equation of state were introduced in the papers [3-5], (see also the examples in [6]). The ideal fluid with a time-dependent non-linear inhomogeneous equation of state has been considered in the papers [7, 8]. Some examples of such kind of equations can be found in papers [9, 10]. It is known that the time-dependence of the parameters in the equation of state of ideal fluid can lead to a transition from the non-phantom era to phantom one in the evolution of the universe (see [11, 12]). The cases where the parameters $w$ and $\Lambda$ in the equations of state are stationary or they depend linearly on time are also described. Under a corresponding choice of the interaction between a dark energy and a dark matter the expressions for a scale factor and its first and second derivatives are obtained. The values of the parameters where the universe expands with acceleration (quintessence) and with deceleration are found. It is shown that the interaction between a dark energy and a dark matter may lead to a periodical appearance of the cosmological singularities [1,13]. Note that time-dependent equation of state fluid may serve as classical analog of string landscape [14] or of modified gravity [15].

## 2. THE INFLUENCE OF THE INTERACTION BETWEEN A DARK ENERGY AND A DARK MATTER ON THE EVOLUTION OF THE UNIVERSE

Let us consider a universe filled with two interacting ideal fluids: dark energy and dark matter. In a spatially flat Friedman—Robertson—Walker metric with a scale factor $a(t)$ the background equations are given by [1]:

$$\begin{cases} \dot{\rho} + 3H(p+\rho) = -Q \\ \dot{\rho}_m + 3H(p_m+\rho_m) = Q \\ \dot{H} = -\dfrac{k^2}{2}(p+\rho+p_m+\rho_m) \end{cases} \quad (1)$$

where $H \equiv \dfrac{\dot{a}}{a}$ is the Hubble rate, and $k^2 = 8\pi G$, with the Newton's gravitational constant $G$; $p, \rho$ and $p_m, \rho_m$ are the pressure and the energy density of a dark energy and a dark matter correspondingly; $Q$ is the interaction term between a dark energy and a dark matter. Here a dot denotes the derivative with respect to cosmic time $t$.

The Friedman's equation for the Hubble rate is given by [1]:

$$H^2 = \dfrac{k^2}{3}(\rho + \rho_m). \quad (2)$$

First of all we will investigate the stationary case where the parameters $w_1, w_2$ in the equations of state and the parameter $\Lambda$ don't depend on time:

$$\begin{aligned} p_m &= w_1 \cdot \rho_m \\ p &= w_2 \cdot \rho + \Lambda + a_1 \cdot H \end{aligned}, \quad (3)$$

where $a_1$ is a some constant.

Taking into account equations (1)—(3) we obtain a gravitational equation of motion for the system "energy-matter":

*Address correspondence to this author at the Tomsk State Pedagogical University, Tomsk, Russia; E-mail: TimoshkinAV@tspu.edu.ru

$$\dot{\bar{\rho}} + \sqrt{3}k\sqrt{\bar{\rho}}\left[(1+w)\bar{\rho} + \Lambda + a_1 H^2\right] = 0, \quad (4)$$

where $\bar{\rho} = \rho + \rho_m$, $w_1 = w_2 = w$.

The solution of the equation (4) looks like:

$$\bar{\rho} = \frac{3\Lambda}{\Phi} \cdot tg^2\left(\frac{1}{2}k\sqrt{\Lambda\Phi} \cdot t + C_1\right), \quad (5)$$

where $\Phi = 3(1+w) + a_1 \cdot k^2$, $C_1$ is an integration constant.

Then we obtain the gravitational equation for the dark matter:

$$\dot{\rho}_m + \sqrt{3}k(1+w)\sqrt{\frac{3\Lambda}{\Phi}} tg\left(\frac{1}{2}\sqrt{\Lambda\Phi} \cdot t + C_1\right)\rho_m = Q \quad (6)$$

We choose the interaction term between a dark energy and a dark matter in the following form:

$$Q = \sin\left(\frac{1}{2}k\sqrt{\Lambda\Phi} \cdot t + C_1\right) \quad (7)$$

In this case the energy density of a dark matter is given by:

$$\rho_m = C_2 \cdot \left[\cos\left(\frac{1}{2}k\sqrt{\Lambda\Phi} \cdot t + C_1\right)\right]^{\frac{6(1+w)}{\Phi}} + \sqrt{\frac{\Phi}{\Lambda}} \cdot \frac{\cos\left(\frac{1}{2}k\sqrt{\Lambda\Phi} \cdot t + C_1\right)}{3k(1+w)\left[1 - \frac{1}{2}\left(1 - \frac{a_1 k^2}{3(1+w)}\right)\right]}, \quad (8)$$

where $C_2$ is an integration constant.

Hubble's rate becomes:

$$H(t) = k\sqrt{\frac{\Lambda}{\Phi}} \cdot tg\left(\frac{1}{2}\sqrt{\Lambda\Phi} \cdot t + C_1\right) \quad (9)$$

It occurs a periodic universe with Big Rip type singularity when

$$t = t_s = \frac{\pi(1+2n) - C_1}{k\sqrt{\Lambda\Phi}}, \quad n \in \mathbb{Z}.$$

The derivative of $H(t)$ is equal to:

$$\dot{H}(t) = \frac{k^2 \Lambda}{2\cos^2\left(\frac{1}{2}\sqrt{\Lambda\Phi} \cdot t + C_1\right)} \quad (10)$$

Thus, the universe expands.

The scale factor is given by the expression:

$$a(t) = \exp\left[\int H(t)dt\right] = \frac{C_2}{\left[\cos^2\left(\frac{1}{2}\sqrt{\Lambda\Phi} \cdot t + C_1\right)\right]^{\frac{1}{\Phi}}} \quad (11)$$

The first and second derivatives of the scale factor are correspondingly given by:

$$\dot{a}(t) = \frac{C_2 \cdot \sin\left(k\sqrt{\Lambda\Phi} \cdot t + C_1\right)}{\left[\cos^2\left(\frac{1}{2}k\sqrt{\Lambda\Phi} \cdot t + C_1\right)\right]^{\frac{\Phi^2 - \Phi + 2}{\Phi^2}}}, \quad (12)$$

$$\ddot{a}(t) = k\sqrt{\Lambda\Phi} \cdot \dot{a}(t) \cdot \left[ctg\left(k\sqrt{\Lambda\Phi} \cdot t + C_1\right)\right] + \frac{\Phi^2 - \Phi + 2}{\Phi^2} \cdot tg\left(\frac{1}{2}k\sqrt{\Lambda\Phi} \cdot t + C_1\right) \quad (13)$$

If $0 < t < \frac{\pi}{2k\sqrt{\Lambda\Phi}}$, both derivatives are positive, the universe expands with acceleration (quintessence), and if $\frac{\pi}{2k\sqrt{\Lambda\Phi}} < t < \frac{\pi}{k\sqrt{\Lambda\Phi}}$, one gets $\dot{a}(t) > 0$ but $\ddot{a}(t) < 0$, the universe expands with deceleration.

Let us suppose that both parameters $w_1, w_2$ and $\Lambda$ depend linearly on time:

$$\begin{aligned} w_1(t) &= w_2(t) = ct + b \\ \Lambda(t) &= dt + e \end{aligned}, \quad (14)$$

where $c, b, d, e$ are some constants. This kind of behaviour may be a consequence of the modification of the gravity (for a review, see [2]).

Then the equation (4) acquires the following form:

$$\dot{\bar{\rho}} + \sqrt{3}k(ct+\theta)\bar{\rho}^{\frac{3}{2}} + \sqrt{3}k(dt+e)\bar{\rho}^{\frac{1}{2}} = 0, \quad (15)$$

where $\theta = b + \frac{\sqrt{3}}{3}a_1 k^2 + 1$.

We shall investigate here, for further simplicity, the case $ce = d\theta$. The solution of the equation (15) looks like:

$$\bar{\rho} = \frac{e}{\theta} tg^2\left[\delta\left(t + \frac{\theta}{c}\right)^2 + C_1\right], \quad (16)$$

where $\delta = \frac{\sqrt{3cd}}{4}$, $C_1$ is an arbitrary constant.

Let's write down the gravitational equation for the dark matter:

$$\dot{\rho}_m + \sqrt{3}k(ct+b+1)\sqrt{\frac{e}{\theta}} tg\left[\delta\left(t + \frac{\theta}{c}\right)^2 + C_1\right]\rho_m = Q \quad (17)$$

If we choose the interaction term in the form:

$$Q = 2\delta\left(t + \frac{\theta}{c}\right) \cdot \sin\left[\delta\left(t + \frac{\theta}{c}\right)^2 + C_1\right], \quad (18)$$

then one obtains the solution of the equation (18) as:

$$\rho_m = C_2 \cos^2\left[\delta\left(t + \frac{\theta}{c}\right)^2 + C_1\right] - \cos\left[\delta\left(t + \frac{\theta}{c}\right)^2 + C_1\right] \quad (19)$$

Huble's rate is equal to:

$$H(t) = \frac{k}{\sqrt{3}}\sqrt{\frac{e}{\theta}} tg\left[\delta\left(t + \frac{\theta}{c}\right)^2 + C_1\right] \quad (20)$$

The periodic universe with finite-time cosmological singularity at $t = t_s = \pm\sqrt{\frac{\pi(1+2n) - C_1}{2\delta}} - \frac{\theta}{c}$, $n \in Z$ occurs. In this case, the energy-density and Hubble parameter simultaneously approach the infinity.

The time derivative of $H(t)$ becomes:

$$\dot{H}(t) = \frac{k^2}{2}\sqrt{d} \cdot \frac{t + \frac{\theta}{c}}{\cos^2\left[\delta\left(t + \frac{\theta}{c}\right)^2 + C_1\right]} \quad (21)$$

If $t > -\frac{\theta}{c}$, then the derivative is positive and the universe is expanding.

The scale factor is given by the expression:

$$a(t) = \exp\left[\int H(t) dt\right] = a_0 \cdot \exp\left[\frac{\mu\left(S\left[\sqrt{\delta}\left(t + \frac{\theta}{c}\right)\right] + C_1 - \frac{1}{2}\right)}{\cos^2\left[\delta\left(t + \frac{\theta}{c}\right)^2 + C_1\right]}\right], \quad (22)$$

where $\mu = \sqrt{\frac{2}{3}\pi\frac{ek}{\theta\sqrt{3cd}}}$, $S\left[\sqrt{\delta}\left(t + \frac{\theta}{c}\right)\right]$ is Frenel's integral, $a_0$ is some constant.

If $t \to +\infty$, then $S\left[\sqrt{\delta}\left(t + \frac{\theta}{c}\right)\right] \to \frac{1}{2}$, $a(t) \to a_0$.

Hence, in this case the scale factor tends to constant and the universe tends to become the Minkowski space. The expansion is stopped, and the universe becomes static.

The first and second derivatives of the scale factor are correspondingly equal to:

$$\dot{a}(t) = a(t) \cdot \mu \cdot tg\left[\delta\left(t + \frac{\theta}{c}\right)^2\right] \cdot$$

$$\left\{1 + 2\delta\left(t + \frac{\theta}{c}\right)\frac{\left(S\left[\sqrt{\delta}\left(t + \frac{\theta}{c}\right)\right] - \frac{1}{2}\right)}{\cos\left[\delta\left(t + \frac{\theta}{c}\right)^2 + C_1\right]}\right\}, \quad (23)$$

$$\ddot{a}(t) = \dot{a}(t) \cdot \left[H(t) + \frac{4\delta\left(t + \frac{\theta}{c}\right)}{\sin 2\delta\left(t + \frac{\theta}{c}\right)^2}\right] + \quad (24)$$

$$2\mu\delta a(t) \cdot tg\delta\left(t + \frac{\theta}{c}\right)^2 \cdot$$

$$\cdot \left[\frac{1 + 2\delta\left(t + \frac{\theta}{c}\right)^2 \cdot tg\delta\left(t + \frac{\theta}{c}\right)^2}{\cos\delta\left(t + \frac{\theta}{c}\right)^2}\right] \cdot$$

$$\left[\left(S\left[\sqrt{\delta}\left(t + \frac{\theta}{c}\right)\right] - \frac{1}{2}\right) + \left(t + \frac{\theta}{c}\right)tg\delta\left(t + \frac{\theta}{c}\right)^2\right].$$

If $-\frac{\theta}{c} < t < \sqrt{\frac{\pi}{2\delta}} - \frac{\theta}{c}$, then both derivatives are positive, the universe expands with acceleration (quintessence). Thus, we constructed the oscillating universe which may appear after the matter-dominated phase [16] before acceleration.

### 3. SUMMARY

In this work we have studied a model of the coupling of a dark energy with a dark matter of special form in which we consider the account of the influence of the interaction between a dark energy and a dark matter on the evolution of the universe. The stationary and non-stationary cases for the parameters $w$ and $\Lambda$ in the equations of state are considered. Unlike a model with a pure dark energy, the presence of the interaction term between a dark energy and a dark matter in the special form in the equations of state, leads to a periodic appearance of Big Rip type singularity. It is possible in the non-stationary case that the universe tends to become the Minkowski space and becomes static in the remote future.

### ACKNOWLEDGEMENTS

We thank professor Sergey Odintsov for very useful discussions and valuable remarks.

The work was supported by the project, LRSS № 4489.2006.02 (Russia).